# On The Effectiveness of Kolmogorov Complexity Estimation to Discriminate Semantic Types

Stephen F. Bush, *Senior Member, IEEE*

Todd Hughes, Ph. D.

**Abstract— We present progress on the experimental validation of a fundamental and universally applicable vulnerability analysis framework that is capable of identifying new types of vulnerabilities before attackers innovate attacks. This new framework proactively identifies system components that are vulnerable based upon their Kolmogorov Complexity estimates and it facilitates prediction of previously unknown vulnerabilities that are likely to be exploited by future attack methods. A tool that utilizes a growing library of complexity estimators is presented. This work is an incremental step towards validation of the concept of complexity-based vulnerability analysis. In particular, results indicate that data types (semantic types) can be identified by estimates of their complexity. Thus, a map of complexity can identify suspicious types, such as executable data embedded within passive data types, without resorting to predefined headers, signatures, or other limiting *a priori* information.**

## I. Introduction

This paper assumes a fundamental understanding of Kolmogorov Complexity details, of which can be found in [13]. Vulnerability analysis as presented in this paper takes into account the innovation of an attacker. A metric for innovation is not new; 700 years ago William of Occam suggested a technique [12]. The salient point of Occam's Razor and complexity-based vulnerability analysis is that the better one understands a phenomenon, the more concisely the phenomenon can be described. This is the essence of the goal of science: to develop theories that require minimum size to be fully described. Ideally, all the knowledge required to describe a phenomenon can be algorithmically contained in formulae;

formulae that are larger than necessary lack a full understanding of the phenomenon. Consider an attacker as a scientist trying to learn more about his environment, that is, the target system. Parasitic computing [1] is a literal example of a scientist studying the operation of a communication network and utilizing it to his advantage in an unintended manner. The attacker as scientist generates hypotheses and theorems. Theorems are attempts to increase understanding of a system by assigning a cause to an event, rather than assuming all events are randomly generated. If theorem $x$, described in bits, is of length $l(x)$, then a theorem of length $l(m)$, where $l(m)$ is much less than $l(x)$, is not only much more compact, but also $2^{l(x)-l(m)}$ times more likely to be the actual cause than pure chance [12]. Thus, the more compactly a theorem can be stated, the more likely one understands the underlying cause. A very literal example of an attacker as a scientist who is trying to understand a phenomenon, i.e. the system being attacked, can be seen in [1]. The results in this paper analyze whether estimates of complexity have the required resolution to differentiate known types of data based upon their complexity.

Early results simulated in Swarm, as well as an active network Java complexity probe toolkit [8][5][3][2][9] had formed the software base and motivating results for this work. A significant result of that work has been the proposition that complexity plays a critical role in security and can be broadly applied as the basis for security analysis and design. Tools based upon this paradigm do not require detailed *a priori* information about known attacks, but rather compute vulnerability based upon an inherent, underlying property of information itself, namely, its Kolmogorov-Chaitin complexity. The author suggests that active networks [3] provide an ideal platform for studying the interaction and properties of algorithmic information and static sequences of data because data can change form deep within, and as it travels through, the network

### A. Innovation and Security

Imagine a vulnerability identification process that consists of the following: waiting for an information system to be attacked, then, assuming it survives and one can detect the attack, analyzing the attack, and if the information system is still not compromised, adding this information to one's knowledge base. This technique would be unacceptable to

Manuscript received Friday, September 26, 2003; intended for submission to the SFI Workshop on Adaptive and Resilient Computing Security. Lockheed Martin Shared Vision funded this work in full.

Stephen F. Bush is with the GE Global Research Center, Niskayuna, NY, 12309, USA phone: 518-387-6827; fax: 518-387-4042; e-mail: bushsf@research.ge.com, http://www.research.ge.com/~bushsf.

Todd Hughes is with Lockheed Martin Advanced Technology Laboratories, Senior Member, Engineering Staff, Lockheed Martin Advanced Technology Laboratories, 3 Executive Campus, Cherry Hill, NJ 08002, Voice: 856.792.9756, Fax: 856.792.9925.

most people, but it is essentially the technique used today. Information assurance, and vulnerability analysis in particular, are hard problems primarily because they involve the application of the scientific method by a defender to determine a means of evaluating and thwarting the scientific method applied by an attacker (see Figure 1 for a conceptual illustration) [6][7].

This self-reference of scientific methods would seem to imply a non-halting cycle of hypothesis and experimental validation being applied by both offensive and defensive entities, each affecting the operation of the other. Information assurance depends upon the ability to discover the relationships governing this cycle and then quantifying and measuring the progress made by both an attacker and defender. A metric and framework are required for quantifying information assurance in such an environment of escalating knowledge and innovation. Progress in vulnerability analysis and information assurance research cannot proceed without fundamental metrics. The metrics should identify and quantify fundamental characteristics of information in order to guarantee assurance.

A fundamental definition of vulnerability analysis is formulated in this paper based upon attacker and defender as reasoning entities, capable of innovation. Truly innovative implementations of attack and defense lead to the evolution of complexity in an information system. Understanding the evolution of complexity in a system enables a better understanding of where to measure and how to quantify vulnerability. The design and implementation of a complexity-based technique is presented in the form of a vulnerability analysis tool for automated measurement of information assurance. The motivation for complexity-based vulnerability analysis comes from the fact that complexity is a fundamental property of information and can be ubiquitously applied to determine vulnerability.

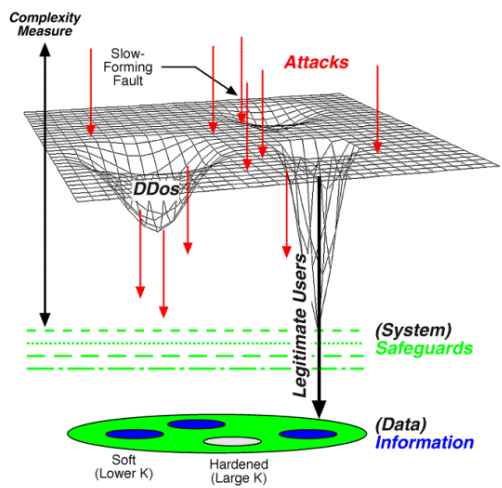

**Figure 1: A Kolmogorov Complexity Map (K-Map) Quantifies Vulnerabilities in a Network.**

## B. Challenges

Complexity-based vulnerability analysis faces many challenges. In particular, the length of time required to obtain an accurate sample (with the goal of performing the analysis in real-time) is critical. In addition, a stream of data on a network link can be sampled at many possible protocol layers. What effect do layers of protocol and encoding have upon the complexity of an information stream? Can these effects be treated as noise and filtered out? The attacker may view protocol and data at a variety of levels: bit, system call, object code, source code, script, etc… All of these levels coexist simultaneously and need to be considered as part of the Kolmogorov Complexity Map of the system. "Map" is a less than perfect term, because given the simultaneous coexistence of layers of protocol and encoding in an information stream, a three dimensional view (that includes complexity, component location, and depth through layers of encoding) might be more apt. A potential attacker is looking for areas of complexity low enough to observe, understand, and thus control (yet that also provide a path to a potential target) but also, in some cases, high enough in complexity in which to potentially hide activity. However, the ability for the attacker to successfully hide his activity depends upon first obtaining a good understanding of the system (defined as a function of complexity) in which to hide. For example, the attacker clearly has a greater understanding of the operation of the rightmost link in Figure 2 based upon its smaller size description of the *complete* operation of the link. By the same measure, the same attacker has a lesser understanding of the operation of the leftmost link. One would expect the attacker to utilize his/her greater knowledge of the rightmost link to spoof or otherwise manipulate that link to his/her advantage.

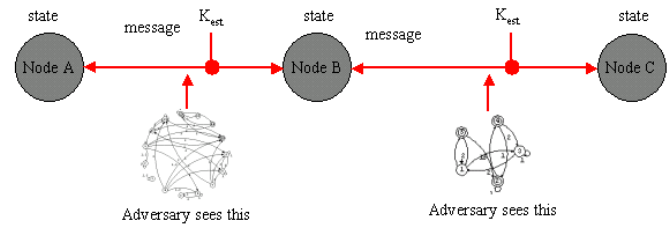

**Figure 2: Attacker's Choice. Less Randomness, Smaller Description Implies Greater Knowledge of Rightmost Link.**

## II. DETECTING VARIATION IN THE COMPLEXITY LANDSCAPE

With regard to generating a complexity map of an information system, a significant assumption is that the complexity landscape has enough variation to allow useful discrimination. It is possible that the smallest descriptive length of different semantic types, depending upon how one chooses to partition the space of semantic types, may be equal or differ by an imperceptible amount. To compound the problem, the best that can be achieved is an approximation of the smallest descriptive length; error is implicit in complexity measurement. One would intuitively expect that the smallest representation of an entity would be the best discriminator because all redundant information is removed and only a unique essence of the entity remains. The techniques used in this paper seek to maximize discrimination capability based upon the smallest representation of a sequence.

An input stream (Figure 3) arriving into the complexity probe's classifier can be comprised of different types of information. The classifier uses a Kolmogorov Complexity estimate of the input stream to categorize incoming data into what is described as a "semantic type": namely, Audio, MS Word Document, Executable, Image, ASCII Text, or Video. Clearly other classification types are possible including finer resolution of types and orientation to other applications, such as genomic information for Bioinformatics applications. The ability to discriminate among semantic types based solely upon the complexity of bit-streams has significant implications. The Kolmogorov Complexity estimate of an actual system can be mapped and such a mapping provides practical and useful information. Detecting embedded semantic types, such as an executable within a document for example, can be performed without signatures or *a priori* information such as tags or "magic numbers" that could be easily spoofed.

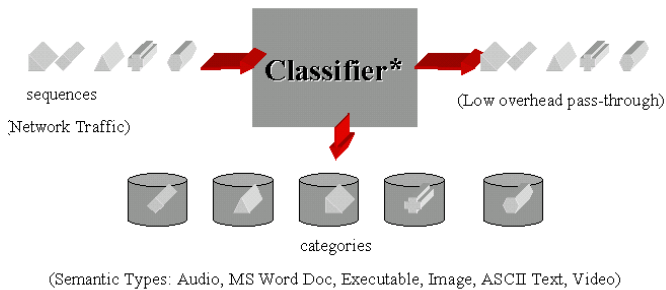

sequences
(Network Traffic)

(Low overhead pass-through)

categories

(Semantic Types: Audio, MS Word Doc, Executable, Image, ASCII Text, Video)

**Figure 3: Experimental Classification of Semantic Types Cased upon Their Complexity.**

*A. Framework and Experimental Test Set*

A consistent experimental test set consisting of at least ten randomly chosen samples of each type of data is used throughout all the experiments in this document. In Figure 4 details of the classifier are shown. Data can be sampled, filtered, and partitioned into windows. The goal of this research has been to avoid sampling and filtering, although components in the code have been implemented to perform this function. Currently, no sampling has been used, that is, all data flows through the system. Filtering has been used only to extract header or payload data when examining protocol data. The complexity estimator examines the input data stream one window at a time and returns an estimate of its complexity. Then the Mapper determines a semantic type based upon the complexity estimate. The data stream can continue to flow through the probe unchanged (although undergoing a delay which is discussed in more detail later in this paper). The semantic type for each window is then returned.

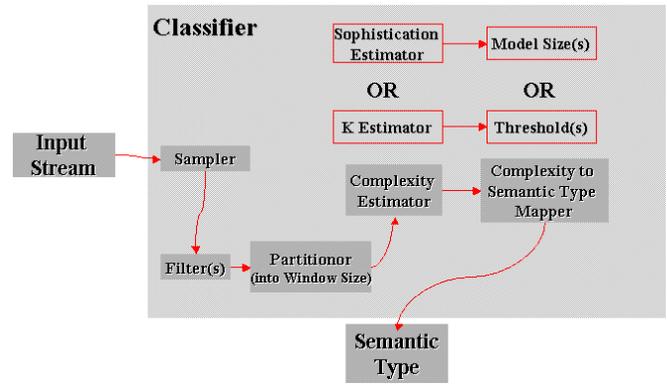

**Figure 4: Components of the Complexity Probe.**

The complexity estimators fit into the complexity estimator module illustrated in Figure 5. The estimator set currently implemented is comprised of a simple entropy estimator (H) whose results are based upon the entropy of a weighting of one bits found in a binary sequence, Limpel-Zev (LZ) compression, Zip (Zip) compression, bZip (bZip) compression, and a frequency-based FFT estimator technique (Psi). The novel Optimal Symbol Compression Ratio (OSCR) technique [10], although shown in the framework, will have its results published in other venues. Note that multiple estimators can work in parallel to discriminate semantic types.

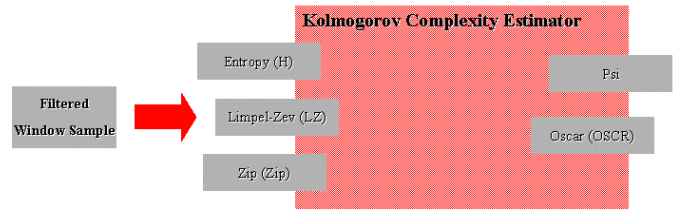

**Figure 5: Plug and Play Complexity Estimators.**

Tunable parameters of the complexity probe are shown in Figure 6, namely, specification of filters, sampling rate, window size, and the set of estimator algorithms enabled. The output can be either a single semantic type to identify a 'file' or a vector of semantic types, one for each window.

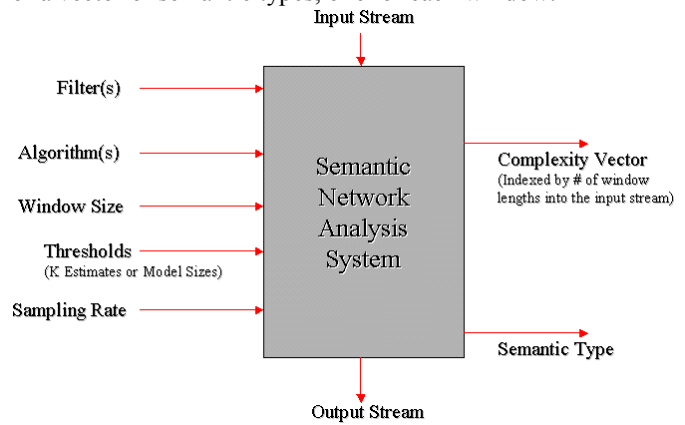

**Figure 6: Tunable Parameters of the Complexity Probe.**

## B. Discrimination Results

Initial techniques used to estimate semantic type based upon complexity used discriminate analysis. The results from a discriminate analysis using only the Zip estimator are shown in Figure 7. The squared distance between semantic types are relatively large except in the case of the distances circled in red. These types are so close to one another that it would be anticipated that these types would yield a high error rate in discriminating among these types. The full paper will elaborate on the efficiency in discriminating these challenging types.

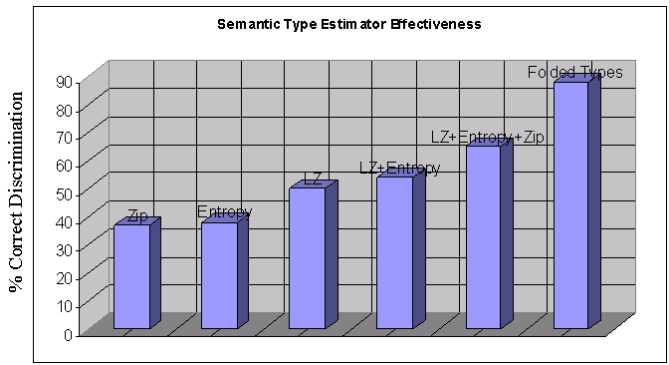

**Figure 7: Discriminant Analysis Estimates of Probe Accuracy.**

Figure 8 shows early results on the accuracy of the complexity-based system to discriminate among semantic types. The histogram columns represent the percent of data from the experimental test set correctly classified. The columns represent Zip, Entropy, Limpel-Zev, Limpel-Zev and Entropy as a combined discriminator, Limpel-Zev and Entropy and Zip as a combined discriminator, and finally the last combined estimator used with audio and executables as a combined type, MS Word and text as a combined type, and Images and video as combined types.

**Figure 8: Early Performance Results for Discrimination of Six Semantic Types.**

## C. Timing Profile

It is hypothesized that the complexity estimator, the actual complexity of the data, and the widow size will have the greatest effects on the timing. Figure 9 shows the mean complexity for each estimator for the entire experimental test set. Zip (Z) yields the lowest expected complexity while Entropy (E) yields the highest. However, in terms of time,

bZip (BZ) and Limpel-Zev (LZ) take the longest expected amount of time with LZ taking the longest.

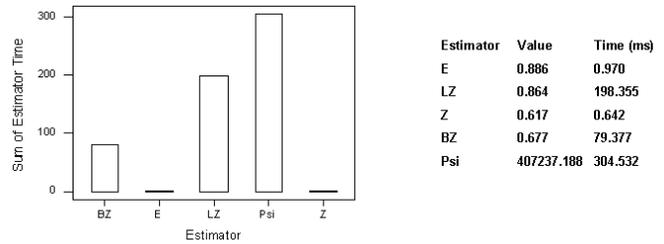

| Estimator | Value | Time (ms) |
|---|---|---|
| E | 0.886 | 0.970 |
| LZ | 0.864 | 198.355 |
| Z | 0.617 | 0.642 |
| BZ | 0.677 | 79.377 |
| Psi | 407237.188 | 304.532 |

**Figure 9: Complexity Estimator Timing Profile.**

In Figure 10 and Figure 11 the expected amount of time for each estimator is shown for each semantic type as a function of both window size and complexity of the sequence in the window. In every case, a larger window size requires more time to estimate complexity. However, a larger window size also yields an increase in throughput. A complete throughput analysis has been performed with selected results shown in Figure 14 and Figure 15.

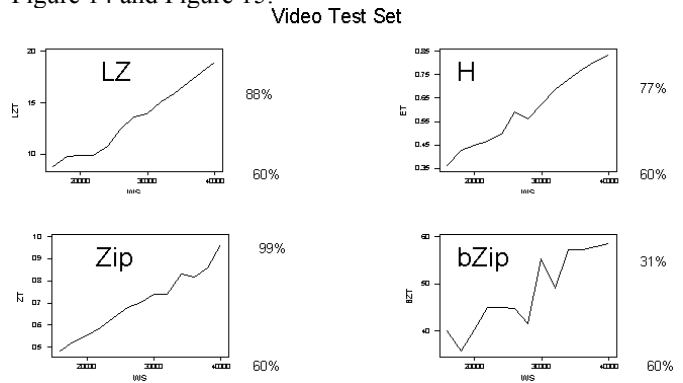

**Figure 10: Time (Milliseconds) Versus Window Size (Bytes) for the Video Data Test Set.**

Clearly the time versus complexity plots show that time to estimate complexity decreases with increasing complexity. This relationship is intuitive since more patterns in a sequence indicate lower complexity and require more time to discern each pattern.

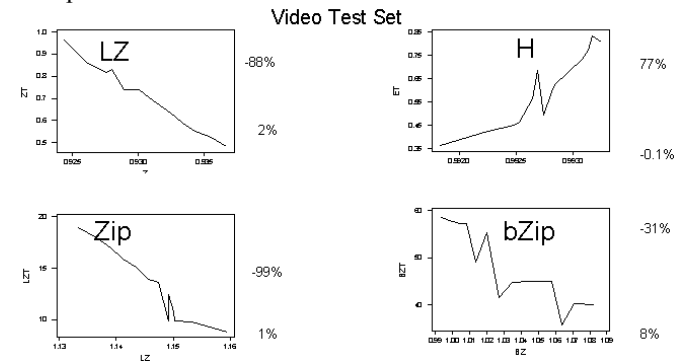

**Figure 11: Time (Milliseconds) Versus Complexity Estimate for Video Data Test Set (Ten Video Files).**

As shown in Figure 12, the measured expected time (milliseconds) for the estimators (and combinations of estimators) to operate upon the complete semantic data set is shown as a function of the percentage of types correctly discriminated. Note that the symbol ('+') in the legend of Figure 12 implies that the corresponding values are used in the process of discrimination, not an explicit summation of the values. Figure 13 illustrates the improvement in discrimination as the compression ratio improves. One interpretation of Figure 13 is that, as a more accurate estimate of complexity is being achieved, discrimination improves.

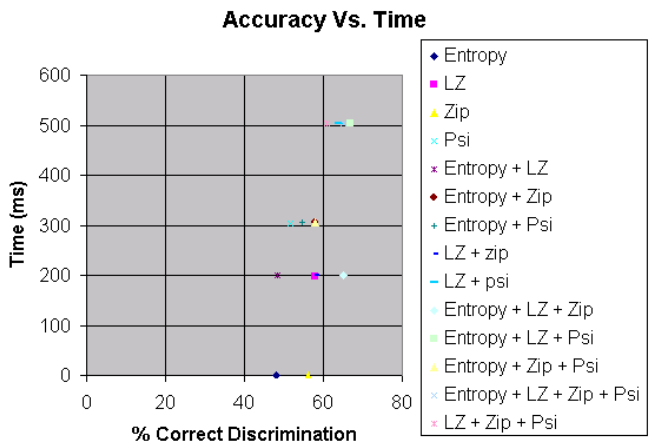

**Figure 12: Timing versus Discrimination Capability.**

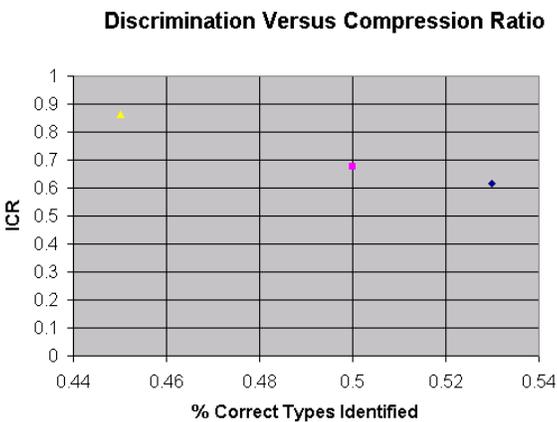

**Figure 13: Better Discrimination with Increased Compression Rates.**

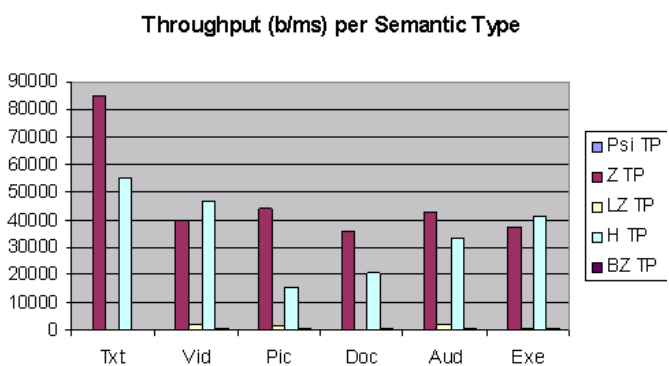

**Figure 14: Throughput for Z and H per Semantic Type.**

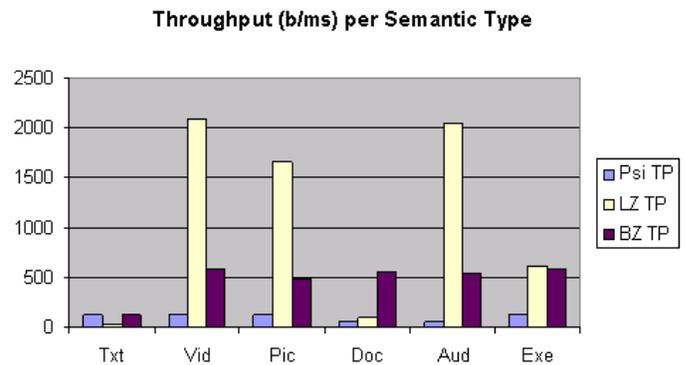

**Figure 15: Throughput for Psi, LZ, and BZ per Semantic Type.**

### III. CONCLUSION

Data types (semantic types) can be identified by estimates of their complexity. Thus, a map of complexity can identify suspicious types, such as executable data embedded within passive data types, without resorting to predefined headers, signatures, or other limiting *a priori* information. Details regarding the mapping of complexity estimates to semantic type and on individual complexity estimators will appear in future papers.

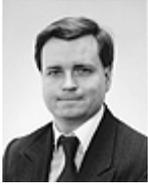

**Stephen F. Bush (M'03-SM'03)** Stephen F. Bush is has over 30 peer-reviewed publications in the areas of communication networking, network modeling and prediction, and complexity theory. Dr. Bush has implemented a toolkit capable of injecting predictive models into an active network. The toolkit has been downloaded and used by more than 600 institutions. Stephen continues to explore novel concepts in complexity and algorithmic information theory to refine the toolkit for applications ranging from network management and ad hoc networking to DNA sequence analyses for bioinformatics applications. Dr. Bush has been the Principal Investigator for many DARPA and Lockheed Martin sponsored research projects including: Active Networking (DARPA/ITO), Information Assurance and Survivability Engineering Tools (DARPA/ISO), and Fault Tolerant Networking (DARPA/ATO). Stephen coauthored a book on active network management, titled Active Networks and Active Network Management: A Proactive Management Framework, published by Kluwer Academic Publishers. Before joining GE Global Research, Stephen was a researcher at the Information and Telecommunications Technologies Center (ITTC) at the University of Kansas. He received his B.S. in Electrical and Computer Engineering from Carnegie Mellon University and M.S. in Computer Science from Cleveland State University. He has worked many years for industry in the areas of computer integrated manufacturing and factory automation and control. Steve received the award of Achievement for Professional Initiative and Performance for his work as Technical Project Leader at GE Information Systems in the areas of network management and control while pursuing his Ph.D. at Case Western Reserve University. Steve completed his Ph.D. research at the University of Kansas where he received a Strobel Scholarship Award. Stephen is currently a Computer Scientist at General Electric Global Research in Niskayuna, NY.

**Todd Hughes, Ph.D.** has more than ten years of experience in ontology as well as knowledge-based approaches to information assurance. He is currently a Senior Member of the Engineering Staff at Lockheed Martin Advanced Technology Laboratories (ATL). Dr. Hughes has served as Principal Investigator on numerous projects including Dynamic Trust-based Resources (DARPA/ATO), Network Mission Assurance (ATL IR&D), and Network Semantic Analysis (ATL-GE Shared Vision IR&D). Prior to joining ATL, Dr. Hughes was an ontological engineer at Cycorp, where he conceived, designed, and developed a Cyc-based network vulnerability assessment application.